\documentclass[english,pra,aps,amssymb,superscriptaddress,showpacs]{revtex4-1}
\usepackage[T1]{fontenc}
\usepackage{amsmath}
\usepackage{amssymb}
\usepackage{graphicx}

\makeatletter
\@ifundefined{textcolor}{}
{%
 \definecolor{BLACK}{gray}{0}
 \definecolor{WHITE}{gray}{1}
 \definecolor{RED}{rgb}{1,0,0}
 \definecolor{GREEN}{rgb}{0,1,0}
 \definecolor{BLUE}{rgb}{0,0,1}
 \definecolor{CYAN}{cmyk}{1,0,0,0}
 \definecolor{MAGENTA}{cmyk}{0,1,0,0}
 \definecolor{YELLOW}{cmyk}{0,0,1,0}
}

\usepackage{epsfig}
\usepackage{dcolumn}
\usepackage{bm}

\makeatother

\usepackage{babel}
\begin{document}

\title{ 
Quench in the 1D Bose-Hubbard model: 
Topological defects and excitations from the Kosterlitz-Thouless phase transition dynamics
}

\author{Jacek Dziarmaga}
\affiliation{Instytut Fizyki Uniwersytetu Jagiello\'nskiego,
             ul. Reymonta 4, PL-30059 Krak\'ow, Poland}

\author{Wojciech H. Zurek}
\affiliation{Theoretical Division, LANL, Los Alamos, New Mexico 87545, USA}

\date{\today}

\begin{abstract}
Kibble-Zurek mechanism (KZM) uses critical scaling to predict density of topological defects and other excitations created in second order phase transitions. We point out that simply inserting asymptotic critical exponents deduced from the immediate vicinity of the critical point to obtain predictions can lead to results that are inconsistent with a more careful KZM analysis based on causality -- on the comparison of the relaxation time of the order parameter with the ``time distance'' from the critical point. As a result, scaling of quench-generated excitations with quench rates can exhibit behavior that is locally (i.e., in the neighborhood of any given quench rate) well approximated by the power law, but with exponents that depend on that rate, and that are quite different from the naive prediction based on the critical exponents relevant for asymptotically long quench times. Kosterlitz-Thouless scaling (that governs e.g. Mott insulator to superfluid transition in the Bose-Hubbard model in one dimension) is investigated as an example of this phenomenon.
\end{abstract}


\maketitle


The study of the dynamics of second-order phase transitions started with the observation by Kibble \cite{Kibble76,Kibble80}
that, in the cosmological setting, as a result of relativistic causality, distinct domains of the nascent Universe will
choose different broken symmetry vacua. Their incompatibility, characterized by the relevant homotopy group,
will typically lead to topological defects that may have observable consequences.

In condensed matter (where the relativistic casual horizon is no longer a useful constraint) one can nevertheless define \cite{Zurek85,Zurek93,Zurek96} a sonic horizon that plays a similar role. The usual approach to estimating the size of the sonic horizon relies on the scaling of the relaxation time and of the healing length that are summed up by the critical exponents. Critical exponents define the universality class of the transition, and this usually enables prediction of the scaling exponent that governs the number of the generated excitations (e.g., the density of topological defects) as a function of the quench timescale $\tau_Q$ for a wide range of quench rates. 

Here we point out that this simple procedure fails in an interesting and unexpected manner for the Kosterlitz-Thouless universality class. That is, one can expect that - in the asymptotic regime where the transition is extremely slow - critical exponents will suffice for such predictions. However, while for the quench rates attainable in the laboratory one may still expect an approximate power law that relates density of excitations to the quench rate, the exponent that characterizes it will begin to approach predictions based on the critical exponents only asymptotically, and for unrealistically (one might even say, astronomically) large values of the ``sonic horizon''. Nevertheless, we show that the application of KZM can lead to predictions that are valid before the asymptotic regime characterized by the critical exponents becomes relevant.

Timescale $\hat t$ at which the ``reflexes'' of the order parameter of the system, quantified by the relaxation time $\tau$, are too slow for its state
to remain in approximate equilibrium with its momentary Hamiltonian (or free energy) controlled from the
outside by the experimenter plays a key role. It is obtained from the equation\cite{Zurek85,Zurek93,Zurek96}:
\begin{equation}
\tau(\epsilon(\hat t))=\epsilon(\hat t)/\dot\epsilon(\hat t)
\label{KZEq}
\end{equation}
that compares relaxation time $\tau$ with the rate of change of the dimensionless distance from the critical point,
e.g. $\epsilon=(T_c-T)/T_c$ where $T_c$ is the critical temperature. When $\epsilon(t)$ is taken to vary on a quench
timescale $\tau_Q$ as
\begin{equation}
\epsilon(t)=t/\tau_Q
\end{equation}
equation (\ref{KZEq}) leads to:
\begin{equation}
\tau(\epsilon(\hat t))=\hat t.
\end{equation}  
In phase transitions where the critical slowing down and critical opalescence can be characterized by power law
dependencies of relaxation time and healing length,
\begin{equation}
\tau(\epsilon)=\tau_0/|\epsilon|^{\nu z},~~~\xi(\epsilon)=\xi_0/|\epsilon|^\nu,
\label{CritEq}
\end{equation}
equation (\ref{KZEq}) can be imidiately solved:
\begin{equation}
\hat t=\tau_0 (\tau_Q/\tau_0)^{\frac{\nu z}{1+\nu z}},~~~\hat\epsilon=(\tau_0/\tau_Q)^{\frac{1}{1+\nu z}}  .
\label{hatepsilon}
\end{equation}
Above, $\nu$ and $z$ are the spatial and dynamical critical exponents that characterize the universality class of the transition, while $\tau_0$ and $\xi_0$ are dimensionful parameters determined by the microphysics. This leads to the characteristic scale
\begin{equation}
\hat\xi=\xi_0(\tau_Q/\tau_0)^{\frac{\nu}{1+\nu z}}.
\label{hatxi}
\end{equation}
It gives the size of the domains that break symmetry in unison, and, hence, dictate the density of topological 
defects left behind by the transition.

Basic tenets of the above Kibble-Zurek mechanism have been confirmed by numerical simulations \cite{LagunaZ1,YZ,DLZ99,ABZ99,ABZ00,BZDA00,ions20,ions2,WDGR11,dkzm1,dkzm2,Nigmatullin11,DSZ12}, and, to 
a lesser degree (and with more caveats) by experiments \cite{Chuang91,Bowick94,Ruutu96,Bauerle96,Carmi00,Monaco02,Monaco09,Maniv03,Sadler06,Anderson08,Golubchik10,Chae12,Griffin12,Schaetz13,EH13,Ulm13,Tanja13,Lamporesi13} in a variety of settings. Refinements include phase transition in inhomogeneous systems (see \cite{DKZ13} for recent overview) and applications of KZM that go beyond topological defect creation (see e.g. \cite{DQZ11,Zurek09,DZ10,Cincio}). Recent reviews related to KZM are also available \cite{Kibble03,Kibble07,Dziarmaga10,Polkovnikov11,DZ13}.

Our aim here is to note that when the critical slowing down is given by a more complicated dependence then the simple power law of Eq. (\ref{CritEq}), the resulting $\hat t$ and, therefore, $\hat\xi$ will vary in a way that cannot be fully characterized by the critical exponents that otherwise suffice to predict their scaling with the quench rate. That is, topological defects or other excitations left behind by the quench will approach the scaling predicted by the critical exponents, Eq. (\ref{hatxi}), only asymptotically, and begin to conform with it only in the regime of extremely slow transitions that may be well out of the reach of laboratory experiments. In the regime of faster quenches that may be accessible to experiments a power law may still be locally a reasonable fit, although its exponent will vary slowly, approaching the asymptotic prediction only very gradually.

\begin{figure*}[t]
\begin{center}\includegraphics[width=15cm]{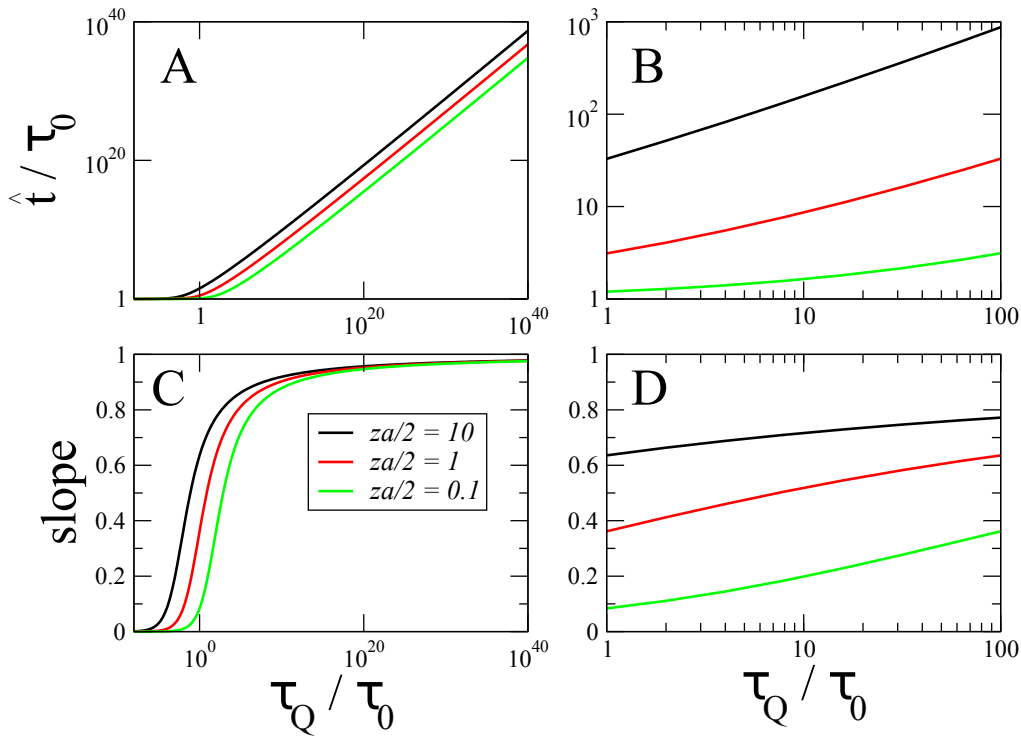}\end{center} 
\vspace{-0.0cm}
\caption{
In the textbook version of the Kibble-Zurek mechanism, the time $\hat t$ when the time evolution ceases to be adiabatic satisfies
a power law $\hat t\propto\tau_Q^{\nu z/(1+\nu z)}$. In a log-log plot this power law becomes a linear function
$\log_{10}\left(\hat t/\tau_0\right)=\frac{\nu z}{1+\nu z}\log_{10}\left(\tau_Q/\tau_0\right)+{\rm const}$, 
where $\tau_0$ is a characteristic timescale of the system. In A, we plot $\hat t$ for a Kosterlitz-Thouless transition 
in function of $\tau_Q$ over many decades of the argument. This function may appear linear locally, i.e., 
in a range of one or two decades, but it actually becomes linear only for very slow quenches, and, consequently, for ``astronomical'' values of the frozen-out domain size $\hat \xi$, Eq. (\ref{hatxi}). 
Indeed, in B, we focus on the narrow range of $\tau_Q=10^{0...2}\tau_0$ that are small enough for a realistic experiment.
These plots may be reasonably approximated by linear functions.
In C, a local slope $d\log_{10}\left(\hat t/\tau_0\right)/d\log_{10}\left(\tau_Q/\tau_0\right)$ of the log-log plot
in panel A in function of $\tau_Q$. The slope $1$, predicted in the critical limit when formally $\nu\to\infty$, is achieved
but only for $\tau_Q$ in the ``astronomical'' regime. When we focus on more realistic $\tau_Q$, as in panel D, the local slope turns out
to be significantly lower than in the critical limit.
}
\label{fig:hatt} 
\end{figure*}

\section*{Results}

{\bf Kibble-Zurek mechanism in the Kosterlitz-Thouless universality class.}
This conclusion about the local power law dependence that approaches scaling dictated by the asymptotic vales of critical exponents is exemplified by the Kosterlitz-Thouless (KT) transition \cite{KT1,KT2,KT3}. There the non-polynomial
scaling of the healing length
\begin{equation}
\xi=\xi_0 \exp(a/\sqrt{|\epsilon|}),
\label{xiKT}
\end{equation} 
where $a\simeq 1$, is captured by stating that the spatial critical exponent $\nu=\infty$, see e.g. \cite{Girvin}. 
This is a brief and dramatic way to sum up the faster than polynomial divergence of $\xi$, 
but it may tempt one to misuse Eqs. (\ref{hatepsilon},\ref{hatxi}).
Thus, formally, one could insert $\nu=\infty$ relevant for the KT universality class into Eq. (\ref{hatxi}) to obtain:
\begin{equation}
\hat\xi=\xi_0\left(\tau_Q/\tau_0\right)^{1/z}.
\label{hatxinaive}
\end{equation}
This equation may be (as we shall see below) asymptotically valid, but is unlikely to have the same range of validity as Eqs. (\ref{hatepsilon},\ref{hatxi})
regarded as the consequence of Eq. (\ref{KZEq}). In particular, for large $\tau_Q$ the exponent $\frac{\nu}{1+z\nu}$ approaches $1/z$, reflected in 
Eq. (\ref{hatxinaive}), only gradually. 

To see why, consider the equation for the relaxation time $\tau\propto\xi^z$ in the KT universality class:
\begin{equation} 
\tau(\epsilon)=\tau_0\exp\left(za/\sqrt{|\epsilon|}\right)
\label{tauKT}
\end{equation}
and assume, as before, $\epsilon=t/\tau_Q$. Equations (\ref{KZEq}) and (\ref{tauKT}) yield
\begin{equation}
\tau_0\exp\left(za/\sqrt{\hat t/\tau_Q}\right)=\hat t.
\end{equation} 
Thus, $\hat t$ is now a solution of a transcendental equation. It can be obtained as
\begin{equation} 
\hat t/\tau_0=(\tau_Q/\tau_0)\left(\frac{\left(za/2\right)}{W\left[\left(za/2\right)\sqrt{\tau_Q/\tau_0}\right]}\right)^2,
\end{equation}
where $W$ is the Lambert function. The above solution is plotted for different values of $za/2$ in Fig. \ref{fig:hatt}A,B.
This relation has been derived before and tested by numerical simulations in a 2D classical model in Ref. \cite{BKTclass}.

Figure \ref{fig:hatt}C shows that the slope of unity for the dependence of $\hat t$ on $\tau_Q$ (and, therefore, $\hat\xi$ on $\tau_Q^{1/z}$) 
is attained only for $\tau_Q$ many orders of magnitude larger than $\tau_0$ -- for exceedingly slow quenches that are unlikely to be 
experimentally accessible. For even reasonably slow quenches the effective power law would be significantly less than $1$, typically as small 
as $\sim0.5$ for $\tau_Q\sim 10\tau_0$, gradually increasing to $0.8...0.9$ as $\tau_Q/\tau_0$ grows to $\sim 10^{10}$ or so.

Therefore, in transitions that exhibit KT-like non-polynomial scalings and result in symmetry breaking the asymptotic behavior one might have inferred from the critical exponents sets in only in the regime that appears to be out of reach of experiments. For instance, the system would have to be large compared to the $\hat \xi \sim 10^{10} \xi_0$, which means (when we take modest $\xi_0 \sim 10^{-10}$m) that the size of the homogeneous system undergoing the transition should be large compared to $\hat \xi$, say $\sim 10^3 \hat \xi$, or, in other words, kilometers!

\begin{figure*}[t]
\vspace{-1.0cm}
\begin{center}\includegraphics[width=13cm]{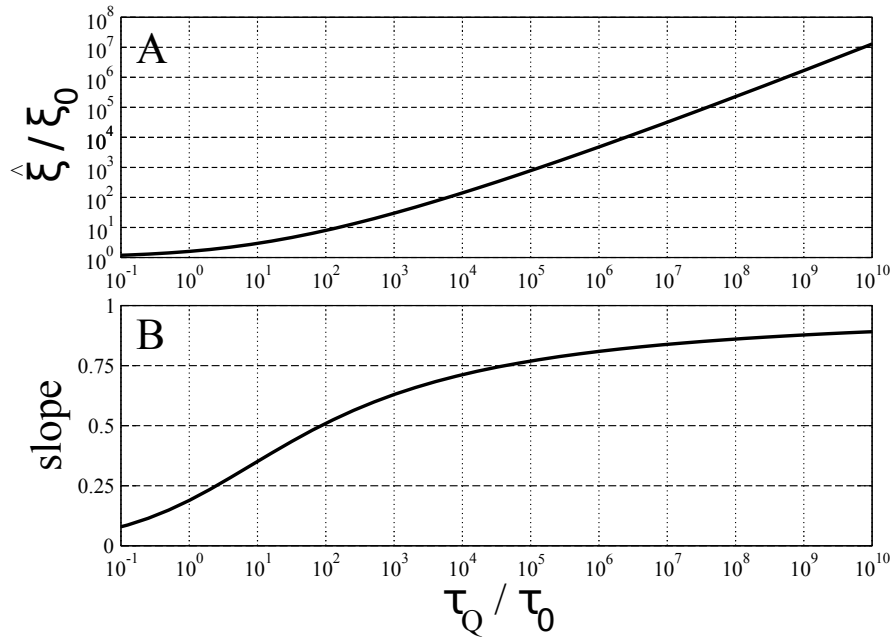}\end{center} 
\vspace{-0.0cm}
\caption{
In A, a log-log plot of the correlation length $\hat\xi$ in function of the quench time $\tau_Q$.
In the textbook Kibble-Zurek mechanism there is a power law $\hat\xi\propto\tau_Q^{\nu/(1+z\nu)}$. 
In a log-log scale this power law would look like a linear function: 
$\log_{10}\left(\hat\xi/\xi_0\right)=\frac{\nu}{1+z\nu}\log_{10}\left(\tau_Q/\tau_0\right)+{\rm const}$.
Our non-linear log-log plot can be reasonably approximated by a linear function locally, i.e., 
over a range of one or two orders of magnitude,
but a local slope of this linearized approximation depends on the order of magnitude of $\tau_Q$,
as shown in panel B. 
Fig. B shows the local slope $d\log_{10}\left(\hat \xi/\xi_0\right)/d\log_{10}\left(\tau_Q/\tau_0\right)$ 
of the log-log plot in panel A in function of $\tau_Q$. For $\tau_Q\to\infty$ the slope tends to $1$, 
as predicted in the critical limit, but for any $\tau_Q$ that is reasonable experimentally it is
significantly less than $1$. For instance, the slope $0.9$ is eventually reached 
at the ``astronomical'' $\tau_Q\simeq 10^{10}\tau_0$,
but for a reasonable $\tau_Q=10^{0...2}\tau_0$ the slope drops to a mere $0.2...0.5$.
}
\label{fig:xi} 
\end{figure*}

A similar difference between the critical limit and the critical regime, although with less dramatic consequences,
arises near the para-to-ferro transition in the random Ising chain \cite{random1,random2,random3}:
\begin{equation} 
H=-\sum_l J_l \sigma^z_l \sigma_{l+1}^z -\sum_l h_l \sigma^x_l,
\end{equation}   
where $J_l$ and $h_l$ are randomly chosen ferromagnetic couplings and transverse magnetic fields respectively.
Here in turn $\nu=2$ is a solid number and it is the dynamical exponent that diverges in the critical regime \cite{randomFisher}: 
\begin{equation}
z=\frac{1}{2|\epsilon|}.
\label{zrandom}
\end{equation}
The limit $\epsilon\to0$, where $z\to\infty$, implies $\hat\xi\sim\tau_Q^0$, i.e., 
a correlation length that does not depend on the quench time at all. 
However, a more careful analysis of the equation (\ref{KZEq}), employing the full formula (\ref{zrandom}) instead of just its critical limit, 
leads to a prediction that there is actually a slow logarithmic dependence on $\tau_Q$, 
a conclusion that was confirmed by simulations in Refs. \cite{random1,random2,random3}. 

A similar care proves beneficial for a non-linear quench
\begin{equation}
\epsilon(t)=\left(\frac{|t|}{\tau_Q}\right)^r {\rm sign}(t)
\label{nonlinear}
\end{equation}
first considered e.g. in Ref. \cite{nonlinear}. Here ${\rm sign}$ is the sign function and $r>0$ is an exponent.
Equation (\ref{KZEq}) yields
\begin{equation}
\hat t\simeq\tau_0 (\tau_Q/\tau_0)^{\frac{\nu z}{1+\nu z}},~~~\hat\epsilon\simeq(\tau_0/\tau_Q)^{\frac{r}{1+\nu z}},
\label{hatepsilon}
\end{equation}
and the characteristic scale of length
\begin{equation}
\hat\xi\simeq\xi_0(\tau_Q/\tau_0)^{\frac{r\nu}{1+\nu z}}.
\label{hatxi}
\end{equation}
Again, this simple but careful argument leads to the same conclusion as the calculations in Ref. \cite{nonlinear}. 

{\bf Kibble-Zurek mechanism in the Bose-Hubbard model. }
We emphasize that KT scaling is encountered in systems other than the classic KT transition in two dimensions (in which generation of vortex pairs occurs via thermal activation as the system is heated). Thus, while for the sake of definiteness, the discussion above was in the framework of finite temperature phase transitions, the universal character
of the arguments makes the conclusions applicable also to quantum phase transitions in the ground state at zero temperature. The most
celebrated example of the quantum KT universality class with $z=1$ is the 1D Bose-Hubbard model \cite{BH}:
\begin{equation}
H=
-J \sum_l \left( b^\dag_{l+1} b_l + b_l^\dag b_{l+1} \right)+
\frac12U \sum_l n_l(n_l-1),
\label{HBH}
\end{equation}
where $b_l$ is a bosonic annihilation operator at site $l$ and $n_l=b_l^\dag b_l$ is an occupation number operator.
At a commensurate filling of $1$ particle per site, the ground state of the model undergoes a K-T quantum phase transition
from a localized Mott phase at $J<J_c$ to a superfluid phase at $J>J_c$. The energy gap on the Mott side of the transition is
\begin{equation}
\Delta=\Delta_0\exp\left(-\frac{a}{\sqrt{x_c-x}}\right).
\end{equation}
Here $x=J/U$ is a dimensionless ratio of the hopping rate $J$ to the interaction strength $U$. 
Using Ref. \cite{Monien2}, it is possible to estimate: $x_c=0.26$, $\Delta_0=0.2J$, and $a=0.3$. 

Any quench from the Mott to the superfluid phase can be linearized near the phase transition
\begin{equation} 
\epsilon(t)=\frac{x_c-x}{x_c}=-t/\tau_Q.
\end{equation} 
The evolution ceases to be adiabatic at $t=-\hat t$ when the reaction time $\Delta^{-1}$ of the system equals 
the time remaining to the transition $|t|$:
\begin{equation}
\exp\left(\frac{a}{\sqrt{x_c\epsilon}}\right)=\Delta_0\hat{t}.
\end{equation}
Its solution is
\begin{equation}
\Delta_0/\hat \Delta=
\frac{\left(\tau_Q/\tau_0\right)}{W\left[\sqrt{\tau_Q/\tau_0}\right]^2},
\end{equation}
where the characteristic timescale is
\begin{equation} 
\tau_0=\frac{4x_c}{a^2\Delta_0}.
\end{equation}
This inverse gap is proportional to the correlation length frozen into the state of the system at $-\hat t$:
\begin{equation} 
\hat\xi \simeq 
\xi_0\left(\Delta_0/\hat \Delta\right)^z =
\xi_0\left(\Delta_0/\hat \Delta\right).
\end{equation}
This correlation length is plotted in Figure \ref{fig:xi}.

To summarize, the equation (\ref{xiKT}) applies in the critical regime where $\epsilon\ll1$ and not only
in the limit $\epsilon\to0$. When the last limit is taken in, say, the Bose-Hubbard model, then the equation 
implies a steep power law $\hat\xi\sim\tau_Q^1$, but a careful application of Eq. (\ref{xiKT}) in the whole
critical regime shows that the steep power law is reached only for rather ``astronomical'' values of
$\tau_Q$ and, especially, of $\hat \xi$ that can hardly be achieved in a realistic experiment. For more realistic quench times 
there is no power law, although in a narrow range of $\tau_Q$ there may appear to be one but with
a much reduced exponent.

\section*{Discussion}

We have seen that, in some cases, using KZM requires more than just inserting critical exponents (that are valid only asymptotically close to the critical point). Rather, to estimate the scale $\hat \xi$ one must make sure that the key idea behind KZM -- critical slowing down as well as the resulting scaling of the sonic horizon -- are accurately described by the critical exponents in the regime probed by the experiment. This may seem like a straightforward requirement, but, as we have seen, there are situations where it may not be easy to satisfy. 

The example with Kosterlitz-Thouless scaling we have just discussed may be extreme in that the scaling represented by the asymptotic values of critical exponents is attained only in the limit that is -- FAPP -- unreachable in the laboratory. Nevertheless, the KZM-like analysis based on the actual dependence of the gap on $ \epsilon$ enables prediction of the scaling modified to suit the range of the experimentally implementable quench rates. 

Key quantity for such considerations is $\hat \epsilon$, the point where the behavior of the system changes character, and the corresponding $\hat t$ that defines the sonic horizon. However, even before one evaluates such subtleties exemplified by the KT transition, it is useful to verify the KZM prerequisite, i.e., whether transition starts and ends sufficiently far from the critical point to justify appeal to KZM. In experiments that involve emulation of condensed matter systems using e.g. trapped ions, or BEC's and optical lattices this may be far from straightforward, as experimental constraints may force relatively short quench timescales (i.e., modest values of $\tau_Q/\tau_0$) which means that $\hat \epsilon$ may be too large -- sonic horizon may be defined too far from the critical point -- to expect near-critical scalings to be relevant. Similar remark applies to sizes of systems: Unless sonic horizon $\sim \hat \xi$ is small compared to the size of the system, scalings predicted by homogeneous KZM will not apply (although -- given certain additional assumptions -- one may be able to deduce their modified versions \cite{DKZ13}). 

A related and interesting issue is how does KZM fail when the assumptions are only approximately satisfied or even violated. Experiments such as \cite{experiment} suggest that this might be a ``soft failure'', i.e., some features of KZM (e.g., power law dependences) may still apply even while detailed predictions (exponents of these power laws) are unlikely to hold. 

There are also indications that even when the requirement of starting and ending the quench on the outside of the $[-\hat \epsilon, +\hat \epsilon]$ interval is satisfied only on one side, KZM like scaling may still emerge. While this is beyond the scope of the original KZM assumptions, it is clearly worthy of a more detailed investigation. 

Indeed, the Bose-Hubbard model is ``gapless'' on the superfluid side, so in this sense only the $-\hat \epsilon$ on the Mott insulator side is well defined. Yet, recent experiment suggests that \cite{experiment} that power laws may approximate the post-quench state of the system, although (at variance with KZM) their slopes appear to depend on where the system starts and ends the quench. Given that the investigated quench times were short ($\tau_Q \simeq \tau_0$), so that quenches likely started and/or ended inside the $[-\hat \epsilon, +\hat \epsilon]$ interval, this is no surprise.

One further complication that is worth noting is that the ``original'' KZM was focused on predicting densities of topologically protected objects. More recent extensions use it to predict other properties of the system following continuous phase transitions. Thus, the size of the sonic horizon has been used to estimate coherence length in the post-transition Bose-Hubbard superfluid. This is, again, an interesting extension, and there are settings (e.g., quantum Ising) where the numerical results (e.g., behavior of entanglement entropy \cite{Cincio}) has been observed. However, as one moves away from the stable and well defined topological defects in integrable systems to less well defined and less stable characteristics (like coherence length in Bose-Hubbard systems that exhibit more complicated behavior), KZM may still yield useful ``guidelines'', but regarding it as prediction without further justification requires courage.

\acknowledgments
This work was supported by the Polish National Science Center (NCN) under Project DEC-2013/09/B/ST3/01603
and by Department of Energy under the Los Alamos National Laboratory LDRD Program.

\vspace{0.5cm}
{\bf Author Contribution:}

\vspace{0.5cm}
J.D. and W.H.Z. jointly defined the project. 
J.D. and W.H.Z. contributed to the analysis, the interpretation, and the preparation of the manuscript.

\vspace{0.5cm}
{\bf Additional Information}

\vspace{0.5cm}
{\it Competing Financial Interests:} The authors declare no competing financial interests.

\end{document}